\documentclass[cits]{JINST}
\usepackage{amsmath}
\usepackage{url}
\usepackage{listings}
\usepackage{relsize}

\title{Imagiro: an implementation of Bayesian iterative unfolding for high energy physics}

\author{Benjamin Wynne$^a$\\
\llap{$^a$}School of Physics and Astronomy, University of Edinburgh, James Clerk Maxwell Building, King's Buildings, Mayfield Rd, Edinburgh. EH8 9EZ\\
  E-mail: \email{bwynne@cern.ch}}

\abstract{Unfolding of reconstructed event properties to identify the true features of collider events is a complementary method to the established practice of detector calibration, and is particularly relevant to large, composite particle detectors such as those at the Large Hadron Collider.
The behaviour of the detector is simulated and used to create a mapping between the true properties of events and their reconstructed equivalents.
Unfolding attempts to invert this mapping for use in correcting measurements.

Imagiro is a new software package providing Bayesian iterative unfolding with systematic and statistical error estimation.
The software is designed to simplify the user experience with automatic self-testing and the calculation of optimal parameters.
Methods are provided for loading data and producing plotted results in the widely used ROOT format.}


\begin{document}

\DeclareRobustCommand{\MC}{^\mathrm{\,MC}}
\DeclareRobustCommand{\data}{^\mathrm{\,data}}
\DeclareRobustCommand{\GEANT4}{\textsc{Geant4}}
\DeclareRobustCommand{\result}{\mathrm{result}}
\DeclareRobustCommand{\hypothesis}{\mathrm{hypothesis}}
\DeclareRobustCommand{\upperhat}{_\mathrm{\,upper}}
\DeclareRobustCommand{\lowerhat}{_\mathrm{\,lower}}
\DeclareRobustCommand{\Npriors}{N_\mathrm{\,priors}}
\DeclareRobustCommand{\Npseudo}{N_\mathrm{\,pseudo-experiments}}
\DeclareRobustCommand{\Nunfolded}{N_\mathrm{\,unfolded}}
\DeclareRobustCommand{\Ndof}{N_\mathrm{\,df}}
\DeclareRobustCommand{\SystematicOf}[3]{ \mathrm{sys}#3\!\big( #1 \vphantom{#1 #2} \big) }
\DeclareRobustCommand{\SigmaOf}[2]{ \sigma\!\big( #1 \vphantom{#1 #2} \big) }
\DeclareRobustCommand{\NumberOf}[2]{ n\!\big( #1 \vphantom{#1 #2} \big) }
\DeclareRobustCommand{\ProbabilityOf}[2]{ P\!\big( #1 \vphantom{#1 #2} \big) }
\DeclareRobustCommand{\GivenProbOf}[2]{ \ProbabilityOf{ #1 | #2 }{#2} }
\DeclareRobustCommand{\InlineCode}[1]{\texttt{\relscale{0.9}{#1}}}

\section{Introduction}
\label{sec:Introduction}

\noindent
Particle detectors in high energy physics (HEP) typically consist of several sub-detectors, each using different methods to make their measurements.
For example, the ATLAS experiment uses silicon pixel and strip trackers, straw tube trackers, lead/liquid argon calorimeters, steel/polystyrene calorimeters and many other technologies~\cite{Aad:1129811}.
Different regions of the detector may use different methods to measure the same quantity, and components in some regions may be disabled due to malfunction.
A particle travelling through the detector may encounter the support structures, power and data cables, cooling pipes and so on that are required to run the detector.
One particle in the detector may also produce many others as it interacts with the material along its path.

In short, the properties of an event may be modified significantly by the detector, and the measurement of these properties is strongly dependent on which area of the detector is involved.

Insofar as possible, these effects are corrected by detailed first-principles treatments of individually understood detector effects such as track-finding efficiencies, hadronic jet scale calibration and muon system punch-through.
However, attempting to correct for every effect this way may be impractical, and runs the risk that some unexpected effect is neglected.
Unfolding can be used as the final step to produce a fully faithful correction of collider signatures to the particle level. 
The principle is that it is easier to calculate the output from the detector given the true properties of an event than to do the reverse.
By simulating physics events in the detector, then simulating the detector itself, it is possible to map the true properties of an event onto those reported by the detector.
Unfolding applies this mapping in reverse to measurements made by the detector in order to recover the true values of observables.

\section{Mapping and unfolding}
\label{sec:Mapping}

\noindent
The simulations of the physics processes under investigation are performed by Monte Carlo generator software such as \textsc{Pythia8}~\cite{Sjostrand:2007gs}, \textsc{Herwig++}~\cite{Bahr:2008pv} and \textsc{Sherpa}~\cite{Gleisberg:2008ta}.
Particle interactions with the detector -- and the creation of secondary particles -- are simulated using a detailed model of the detector hardware in software like \GEANT4~\cite{Agostinelli:2002hh}.
Having simulated the behaviour of particles within the detector, the detector's response is also simulated in a process called `digitisation.'
If \GEANT4 calculates that a particle has encountered detector hardware then the properties of that particle are passed to the corresponding digitisation simulator for that hardware.
While \GEANT4 is software for general use, digitisation simulators are written and tuned specifically to reproduce the electronic response of a particular sub-detector component.
Finally these simulated detector readouts are interpreted in a process called reconstruction, as it attempts to reconstruct the paths and energies of particles from tracking and calorimeter information.
Reconstruction should be identical for digitised events, and for real readout from the detector itself.
Ideally, output from the reconstruction process should match the truth.
In practice this is not the case, and unfolding can account for the discrepancy.

\begin{figure}[H]
	\begin{center}
        \includegraphics[width=0.7\textwidth]{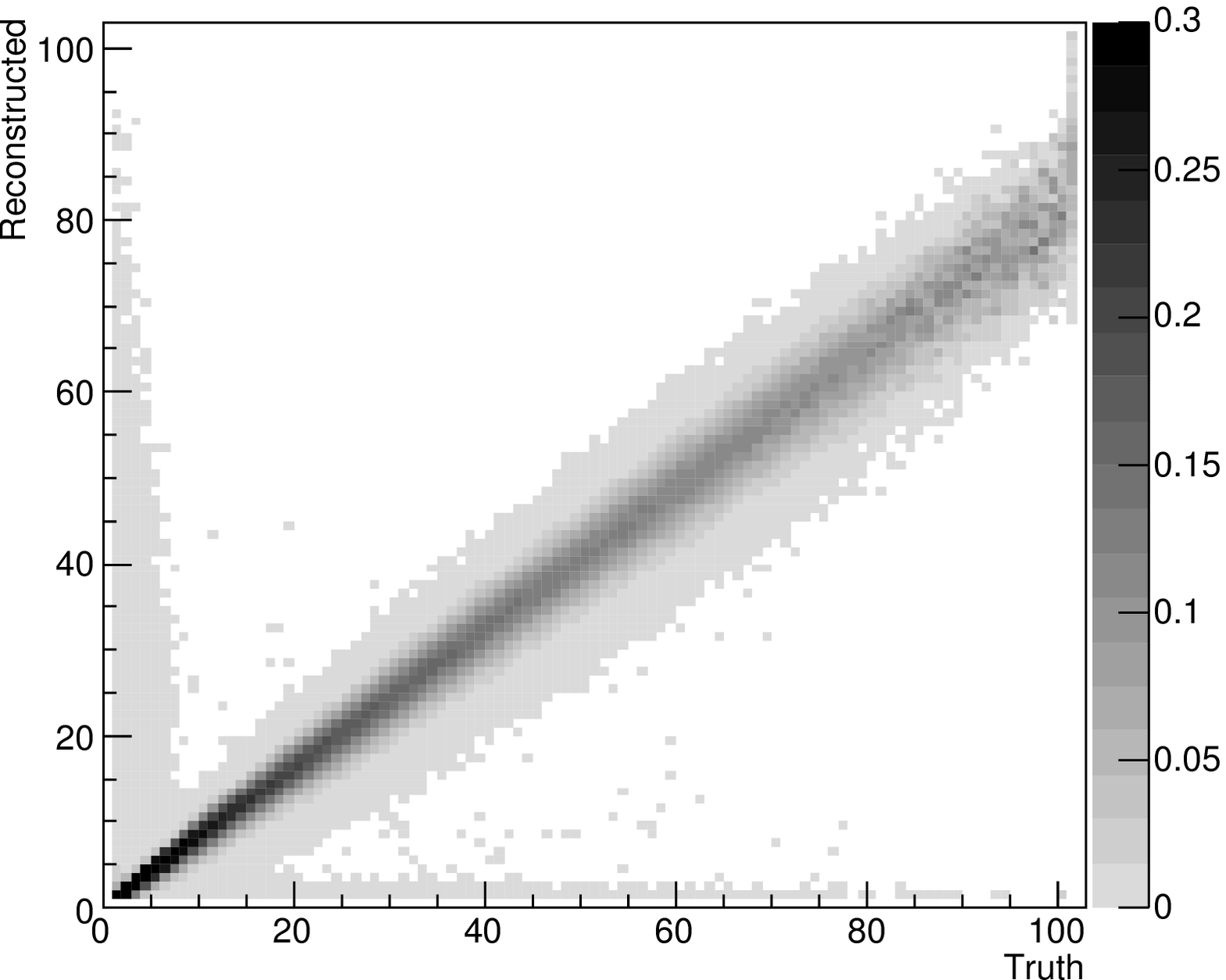}
        \caption[Example smearing matrix.]
        {An example of the smearing matrix -- the mapping between MC truth and reconstructed values of a variable.}
        \label{fig:SmearingMatrix}
	\end{center}
\end{figure}

In the following the original values of observables are referred to as `truth,' and their values as measured by the detector are referred to as `reconstructed.'
Simulated events are referred to as `Monte Carlo' or `MC' and real events as `data.'
Constructing the mapping of truth to reconstructed values is simple if simulated MC events are available: for a given variable one can plot the truth (as reported by the Monte Carlo generator) against the simulated reconstructed value.
Repeating for many truth-reconstructed pairs builds up a two-dimensional histogram of the frequency that a particular MC truth gives a particular reconstructed value: see Fig.~\ref{fig:SmearingMatrix} for an example.
This is only valid for observables with a known connection between the true and reconstructed values, such as the properties of entire events where each event has a unique identifier.
In cases where the connection is more subjective (e.g.\ associating energy in the calorimeter with the energy of specific truth particles) this unfolding technique is not appropriate.

Elements on the diagonal of the truth-to-reconstructed mapping histogram give the probability of a particular MC truth value being correctly reconstructed, while the off-diagonal elements give the probability that a value belonging in bin $i$ is mistakenly measured as belonging in bin $j$.
This mapping is called the `smearing matrix\footnote{Often also referred to as the migration or response matrix.}' $S_{ij}$, and the effect of a detector on the measurement of an observable can be thought of as:
\begin{equation}
        \sum\limits_i S_{ij} T_i = R_j,
	\label{eqn:ApplySmearing}
\end{equation}
where $T_i$ is the true distribution of the observable and $R_j$ is the reconstructed distribution.

Naively, it should be possible to recover the true distribution of a variable by inverting the smearing matrix and applying it to the measured distribution:
\begin{equation}
        \sum\limits_j S_{ij}^{-1} R_j = T_i.
\end{equation}
Unfortunately there is no guarantee that $S_{ij}$ is invertible or that the inversion has a unique solution.
Attempting this method tends to give unphysical results, such as regions of negative probability density in the result distribution, or small changes in input causing disproportionate fluctuations in the output.
Such problems are called `ill-posed' or `improper,' as discussed in references~\cite{IP1972157,Blobel:157405,Hadamard:book}.

Rather than direct inversion of the smearing matrix, D'Agostini proposed an alternative method~\cite{D'Agostini:1994zf}. It uses Bayes' theorem, which is usually stated as:
\begin{equation}
	\GivenProbOf{\hypothesis\,}{\,\result} = \frac{ \GivenProbOf{\result}{\,\hypothesis} \ProbabilityOf{\hypothesis}{\result} }{ \ProbabilityOf{\result}{\hypothesis} },
	\label{eqn:BayesTheorem}
\end{equation}
i.e.\ it relates the probability of a hypothesis being correct given an experimental result to the probability of that hypothesis producing the result.

Restating Eqn.~\ref{eqn:BayesTheorem} in the terminology of unfolding, each hypothesis is a bin $T_i\MC$ in the MC truth distribution histogram and each result a bin $R_j\MC$ in the reconstructed MC histogram:
\begin{equation}
	\GivenProbOf{T_i\MC}{\,R_j\MC} = \frac{ \GivenProbOf{R_j\MC}{\,T_i\MC} \ProbabilityOf{T_i\MC}{R_j\MC} }{ \ProbabilityOf{R_j\MC}{T_i\MC} }.
        \label{eqn:UnfoldingMatrix}
\end{equation}
Now $\GivenProbOf{R_j\MC}{T_i\MC}$ corresponds to the smearing matrix $S_{ij}$ -- a mapping of truth to reconstructed values -- and $\ProbabilityOf{R_j\MC}{T_i\MC}$ is given by
\begin{equation}
	\ProbabilityOf{R_j\MC}{T_i\MC} = \sum\limits_i \GivenProbOf{R_j\MC}{T_i\MC} \ProbabilityOf{T_i\MC}{R_j\MC},
\end{equation}
which is equivalent to Eqn.~\ref{eqn:ApplySmearing}.
We can now use the unfolding matrix $\GivenProbOf{T_i\MC}{R_j\MC}$ from Eqn.~\ref{eqn:UnfoldingMatrix} to correct a measured distribution:
\begin{equation}
	\NumberOf{C_i\data}{R_j\data} = \sum\limits_j \GivenProbOf{T_i\MC}{R_j\MC}\,\NumberOf{R_j\data}{C_i\data},
	\label{eqn:MakeCorrected}
\end{equation}
where $\NumberOf{R_j\data}{C_i\data}$ indicates the number of events in bin $j$ of the measured data histogram, and $\NumberOf{C_i\data}{R_j\data}$ the number in bin $i$ of the corrected histogram.

This method not only requires the smearing matrix as an input, but also a prior distribution $\ProbabilityOf{T_i\MC}{R_j\MC}$.
Using the Monte Carlo truth distribution for the prior seems natural, but raises the question of model-dependence.
Clearly the truth distribution depends on the Monte Carlo generator that created it, and using different priors for the unfolding will give different results.
D'Agostini addresses model-dependence with \emph{iterative} unfolding: take each unfolded distribution and use it as the prior for the next round of unfolding.
Repeating this process should cause the output to converge with the true data distribution.
The convergence of the procedure is discussed in more detail in Section~\ref{subsec:UnfoldingConvergence}.

\section{Imagiro}
\label{sec:UnfoldingImagiro}

\noindent
Imagiro is a software package that provides Bayesian iterative unfolding with robust tests and safeguards.
It attempts to achieve the best unfolding performance automatically, without requiring manual fine-tuning.
The Monte Carlo and data events are loaded from ROOT format~\cite{ROOT} files, and the unfolded distributions stored as ROOT histograms.
Besides the unfolding algorithm there are a number of additional features, described in this section.

\subsection{Multiple priors}
\label{subsec:UnfoldingPriors}

\noindent
Iterating the unfolding procedure should remove the dependence of the result on the prior distribution.
However, this assertion must be tested to gauge the success of any attempted unfolding.
Imagiro is designed so that the whole process can be repeated for multiple different priors (Monte Carlo samples), and the results combined.
The final value for each histogram bin is the mean of the results of unfolding with each prior distribution:
\begin{equation}
	\NumberOf{F_i\data}{C_{i,\,d}\data} = \frac{ 1 }{ \Npriors } \sum\limits_{ d = 1 }^{ \Npriors } \NumberOf{C_{i,\,d}\data}{F_i\data},
	\label{eqn:BinMean}
\end{equation}
where $\NumberOf{F_i\data}{C_{i,\,d}\data}$ is the number of events in bin $i$ of the final result histogram, $\NumberOf{C_{i,\,d}\data}{F_i\data}$ is the number in bin $i$ of the histogram unfolded using prior distribution $d$, and $\Npriors$ is the total number of prior distributions.
Any model dependence is considered a systematic error, as discussed in Section~\ref{subsec:UnfoldingSystematic}.

Ideally, the smearing matrix should depend only on the detector simulation used, and should not depend on the Monte Carlo model.
Therefore if the MC samples use the same detector simulation they can share a single smearing matrix, constructed using all events from all samples.
This reduces the statistical uncertainty, but a separate matrix for each sample can be made instead in order to test the model-independence of the matrix itself.

\subsection{Closure tests}
\label{subsec:UnfoldingClosure}

\noindent
Closure tests are used to demonstrate that the software is working correctly.
Given that the smearing matrix maps the true MC distribution onto reconstructed MC, the software should be able to unfold the reconstructed MC distribution to reproduce MC truth.
By comparing the result from unfolding with the actual MC truth distribution, Imagiro demonstrates that it is performing correctly.
When multiple MC samples are provided, a closure test is performed with each one.
The criterion for success is that $\chi^2 / \Ndof < 1$ in the comparison of the corrected and true distributions, where $\Ndof$ is the number of degrees of freedom.

\subsection{Conditions for convergence}
\label{subsec:UnfoldingConvergence}

\begin{figure}[H]
	\begin{center}
        \includegraphics[width=0.7\textwidth]{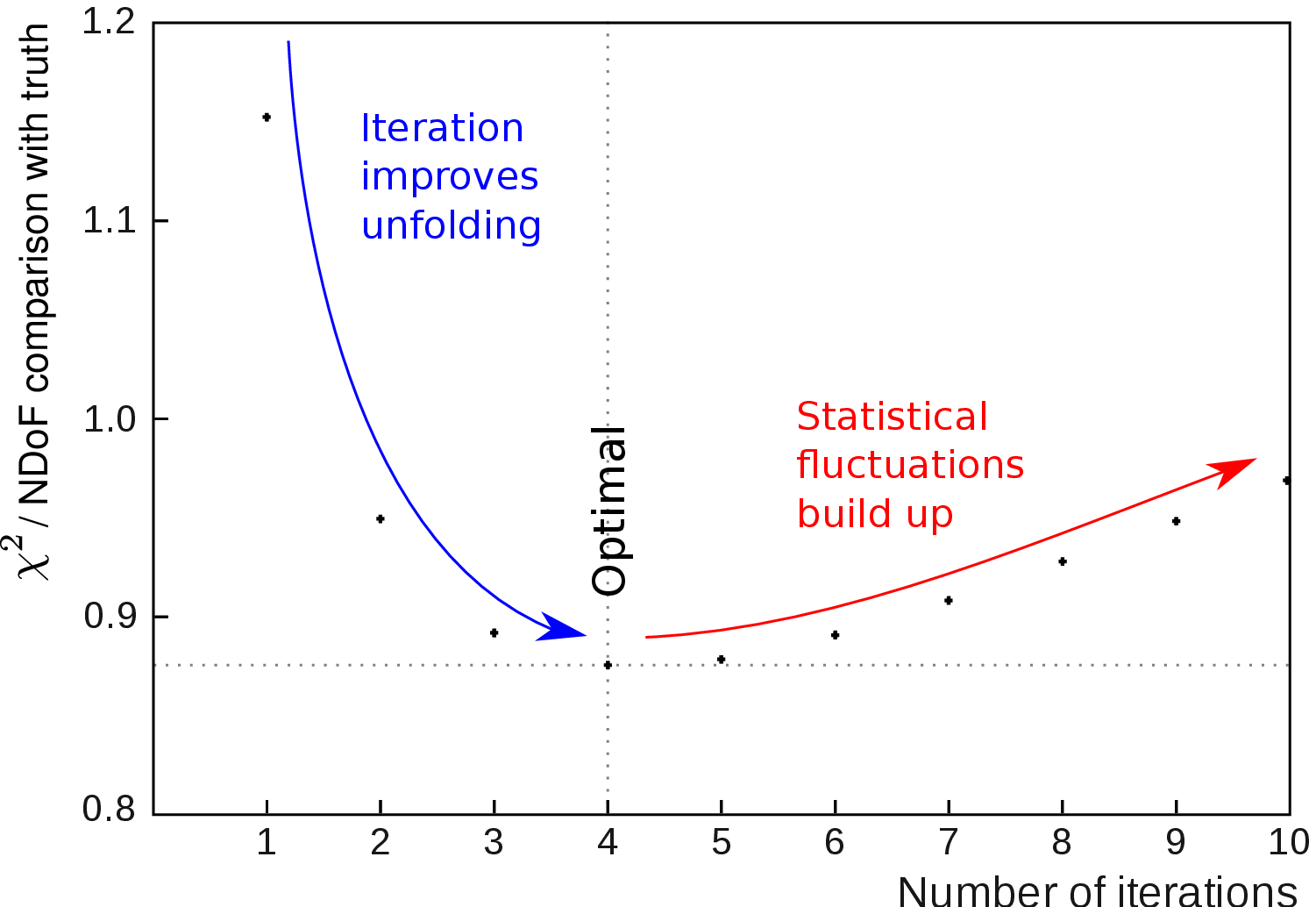}
        \caption[Iterative unfolding converges on the truth.]
        {While the first iterations of unfolding bring the result closer to the truth, there is an optimum point past which the accumulation of statistical errors becomes more significant.
	This is a representative example of results from Imagiro}
        \label{fig:ConvergenceTest}
	\end{center}
\end{figure}

\noindent
Iterating the unfolding process causes the unfolded distribution to converge with the true data distribution.
However, it also compounds the effects of statistical uncertainties in the smearing matrix.
Therefore a larger number of iterations does not guarantee a better result: eventually the true distribution will be obscured by random fluctuations.
There is hence an optimum point when the unfolded distribution most closely describes the truth.
Fig.~\ref{fig:ConvergenceTest} shows an example of this behaviour, by unfolding a reconstructed MC distribution and each iteration comparing the result to the truth.

When multiple MC samples are provided, Imagiro attempts to identify the optimum point automatically.
MC reconstructed distribution $A$ is unfolded with MC truth distribution $B$ as a prior.
Each iteration the output is compared with MC truth $A$, which it should reproduce exactly in an ideal situation.
This process is repeated for all combinations of true and reconstructed distributions from different MC samples, since different priors will give different unfolding performance.
For each pair of MC samples $10$ iterations are performed, and the iteration that gives either the minimum value of $\chi^2 / \Ndof$ or the maximum result of Kolmogorov-Smirnov comparison -- whichever comes first -- is taken to be the optimum.
Taking the mean of the optimal iteration numbers from each of these tests gives the number of iterations to use when unfolding the data.
Note that unfolding a Monte Carlo reconstructed distribution with its own truth as a prior is not attempted because this is equivalent to a closure test.
The unfolding may perform differently for different observables, so the optimum iteration number is calculated separately for each one.

\subsection{Unfolding correlated variables}
\label{subsec:UnfoldingCorrelated}

\noindent
Unfolding corrects distributions, not individual events.
When investigating the correlation between two observables (i.e.\ making a profile histogram) it is not possible to unfold each variable separately, as the correlation information is not preserved in the process.
To preserve these correlations Imagiro first makes a combined observable.
As an example, if one observable has bins $A$, $B$ and $C$, and a correlated observable has bins $x$, $y$, and $z$, then Imagiro would make a combined observable with bins $Ax$, $Ay$, $Az$, $Bx$, $By$, $Bz$, $Cx$, $Cy$ and $Cz$.
Unfolding this combined observable preserves the correlation information.

Converting to and from this combined observable may introduce errors through poor choice of binning.
Good binning of the $y$-axis observable is particularly important, since to construct a profile histogram conventionally only the $x$-axis is binned.
Imagiro checks for mistakes here by constructing two profile histograms from the prior distribution: one without binning the $y$-axis; and one using the same binning as the combined observable.
These two profiles are compared, and if any bin value differs by more than $1$\% the problem is reported.

Potentially this process could be abstracted to higher dimensions, but it is unlikely to be practical.
Already for two correlated observables the smearing matrix histogram must have $( N_{\mathrm{x bins}} \times N_{\mathrm{y bins}} )^2$ total bins, which may require a large number of MC events to populate.
Adding more correlated observables will make the smearing matrix even larger, requiring even more MC events.

\subsection{Fake or missing events}
\label{sec:FakeMiss}

\noindent
It is possible for an MC truth event to have no corresponding partner in the list of reconstructed MC events, or vice versa.
This is likely when events must pass a selection criterion based on uncorrected reconstructed observables.
If a truth event has no reconstructed partner it is called `missed,' and if a reconstructed event has no partner in truth it is called `fake.'
D'Agostini provides an explicit treatment for missed events by introducing an efficiency $\epsilon_i$~\cite{D'Agostini:1994zf}, which modifies Eqn.~\ref{eqn:MakeCorrected} as follows:
\begin{equation}
	\NumberOf{C_i\data}{R_j\data} = \frac{1}{\epsilon_i} \sum\limits_j \GivenProbOf{T_i\MC}{R_j\MC}\,\NumberOf{R_j\data}{C_i\data},
\end{equation}
The efficiency is defined as the probability that an event in bin $i$ of the true distribution has a partner in any bin of the reconstructed distribution.

Imagiro can also correct for fake events by assigning them a prior probability.
Each prior distribution has an extra bin added, and the smearing matrix contains an extra column giving the probability that a particular reconstructed event was fake.

\section{Error treatment}

\noindent
Imagiro provides full statistical and systematic error calculations.
Statistical errors on the MC distributions affect the final result, not just errors from the measured distribution.
Model-dependence of the unfolded results gives a potential systematic error which may not be entirely removed by iteration.
There is also the capability to propagate experimental systematic errors through the unfolding process.
The different error calculations are discussed in this section.

\subsection{Statistical error calculation}
\label{subsec:StatisticalErrors}

\noindent
The calculation of statistical errors on the unfolded distribution is complex, since each bin of the measured distribution and prior distribution has some effect on each bin of the corrected distribution.
Additionally, there is a contribution to the statistical error from the smearing matrix, since that is constructed with a finite number of MC events.
The errors are compounded each iteration as the smearing matrix is re-used.
Imagiro calculates the covariance matrix due to these errors using Adye's correction~\cite{Adye:2011gm} to D'Agostini's original error formula~\cite{D'Agostini:1994zf}.
The calculation can be slow (execution time scales linearly with the number of histogram bins, and combined observables can require many bins), so the option is provided to only calculate the diagonal elements of the covariance matrix.

\subsection{Fast statistical error estimation}
\label{subsec:FastErrors}

\noindent
As an alternative to the full statistical error calculation, a near-instantaneous method is provided.
The statistical error for each bin in the input distribution is scaled proportionately to the correction made to that bin value:
\begin{equation}
	\SigmaOf{C_i\data}{R_i\data} = \SigmaOf{R_i\data}{C_i\data} \frac{ \NumberOf{C_i\data}{R_i\data} }{ \NumberOf{R_i\data}{C_i\data} },
\end{equation}
where $\SigmaOf{C_i\data}{R_i\data}$ is the statistical uncertainty of bin $i$ in the corrected histogram, and $\SigmaOf{R_i\data}{C_i\data}$ is the corresponding error for the measured data.
This method does not account for migrations of events between bins in the unfolding, or for the statistical uncertainty of the smearing matrix, so should be treated with caution.
The covariance matrix is not calculated.

\subsection{Intrinsic systematic error}
\label{subsec:UnfoldingSystematic}

\noindent
Unfolding the measured data with different priors gives different corrected distributions.
Iterating the unfolding reduces the difference between corrected distributions, but it is never eliminated.
This remaining model-dependence is treated as a systematic error intrinsic to the unfolding process.
Each bin of the final result histogram will have $\Npriors$ different unfolded values associated with it.
The central value for that bin is given by the mean of all these unfolded values, as described in Section~\ref{subsec:UnfoldingPriors}.
The systematic error is calculated by ordering all the unfolded values and taking the range of the central $68$\% of them.
If the ordered values for bin $i$ are indexed by $A$ running from smallest to largest, the systematic error is given by:
\begin{align*}
        A\upperhat &= \mathrm{ceiling}( \Npriors \times 0.84 ),\\
	\SystematicOf{F_i\data}{C_{i,\,A\upperhat}\data}{\upperhat} &= \NumberOf{C_{i,\,A\upperhat}\data}{F_i\data} - \NumberOf{F_i\data}{C_{i,\,A\upperhat}\data},
\end{align*}
\begin{align*}
        A\lowerhat &= \mathrm{floor}( \Npriors \times 0.16 ),\\
	\SystematicOf{F_i\data}{C_{i,\,A\lowerhat}\data}{\lowerhat} &= \NumberOf{F_i\data}{C_{i,\,A\lowerhat}\data} - \NumberOf{C_{i,\,A\lowerhat}\data}{F_i\data},
\end{align*}
where $\SystematicOf{F_i\data}{C_{i,\,A\upperhat}\data}{\upperhat}$ and $\SystematicOf{F_i\data}{C_{i,\,A\lowerhat}\data}{\lowerhat}$ are the asymmetric systematic error ranges of bin $i$ in the final result histogram, $\NumberOf{F_i\data}{C_{i,\,A}\data}$ is the number of events in that bin, and $\NumberOf{C_{i,\,A}\data}{F_i\data}$ is the $A^\mathrm{th}$ largest number of events from bin $i$ of each of the different corrected histograms.

\subsection{Systematic error propagation}
\label{subsec:SystematicPropagation}

\noindent
Systematic errors on the measured data distributions can be propagated through the unfolding in Imagiro using pseudo-experiments.
When constructing a data distribution for unfolding, additional pseudo-experiment distributions are made by offsetting the input data values.
The offset for each value is either sampled per-event from a Gaussian with a user-defined width, or is a specific value requested by the user.
Absolute or fractional errors can be specified, and the user can use the value of one event-level observable as the error for another.

Each pseudo-experiment is unfolded with each available prior distribution, just like the measured data.
Therefore each bin of the final result histogram will have $\Npriors \times \Npseudo$ extra unfolded values associated with it, in addition to those from unfolding the measured data without offsets.
The systematic error for that bin is calculated just as in Section~\ref{subsec:UnfoldingSystematic}, except that $A\upperhat$ and $A\lowerhat$ are now given by:
\begin{align*}
	\Nunfolded &= \Npriors ( \Npseudo + 1 ),\\
	A\upperhat &= \mathrm{ceiling}( \Nunfolded \times 0.84 ),\\
	A\lowerhat &= \mathrm{floor}( \Nunfolded \times 0.16 ).
\end{align*}
The final value for bin $i$ is now given by the mean of all $\Nunfolded$ values for that bin, modifying Eqn.~\ref{eqn:BinMean} as follows:
\begin{equation}
        \NumberOf{F_i\data}{C_{i,\,A}\data} = \frac{ 1 }{ \Nunfolded } \sum\limits_{ A = 1 }^{ \Nunfolded } \NumberOf{C_{i,\,A}\data}{F_i\data}.
\end{equation}

\section{Demonstration with toy models}
\label{sec:Demonstration}

\begin{figure}[tp]
	\begin{center}
        \includegraphics[width=0.7\textwidth]{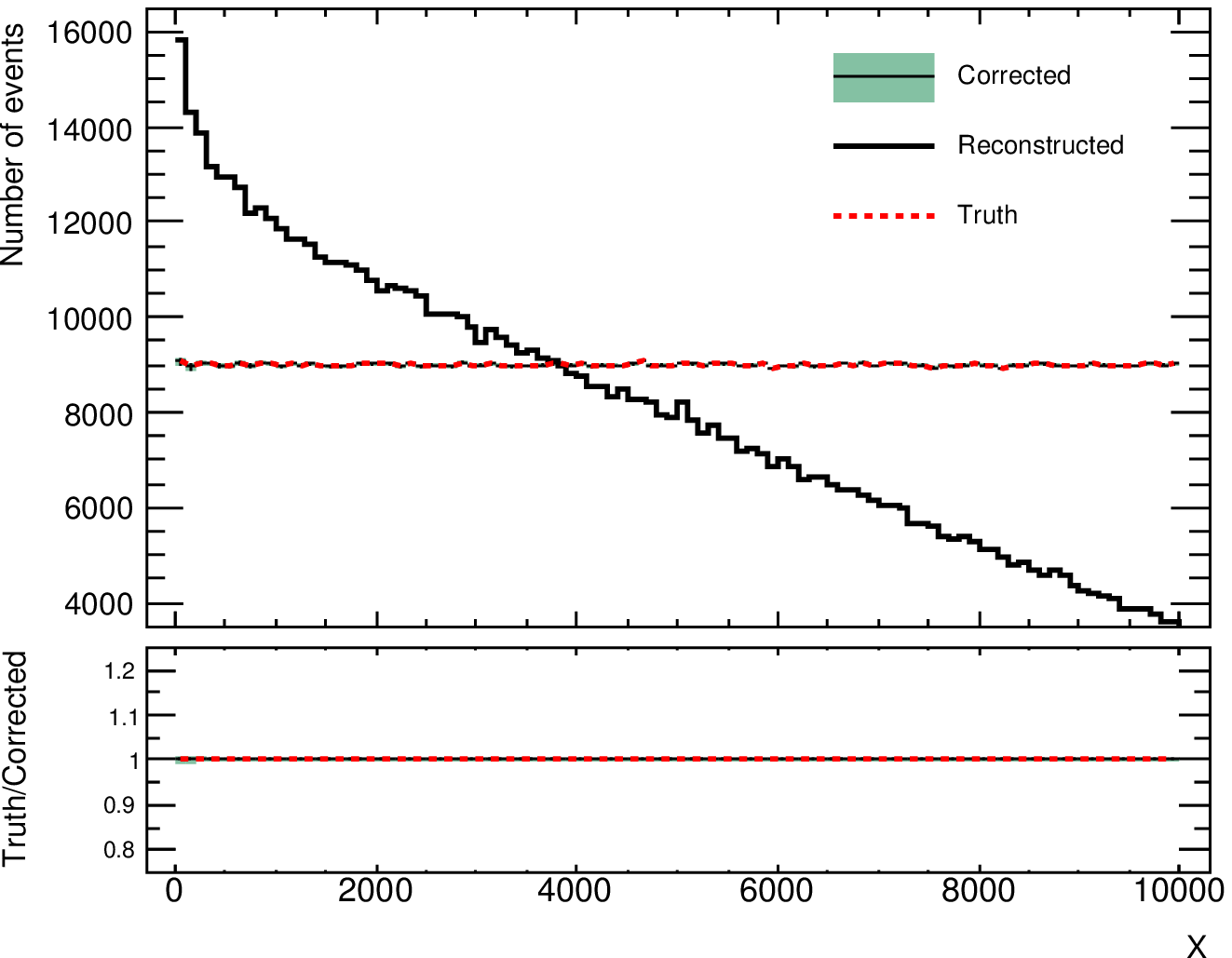}\\[1em]
        \includegraphics[width=0.7\textwidth]{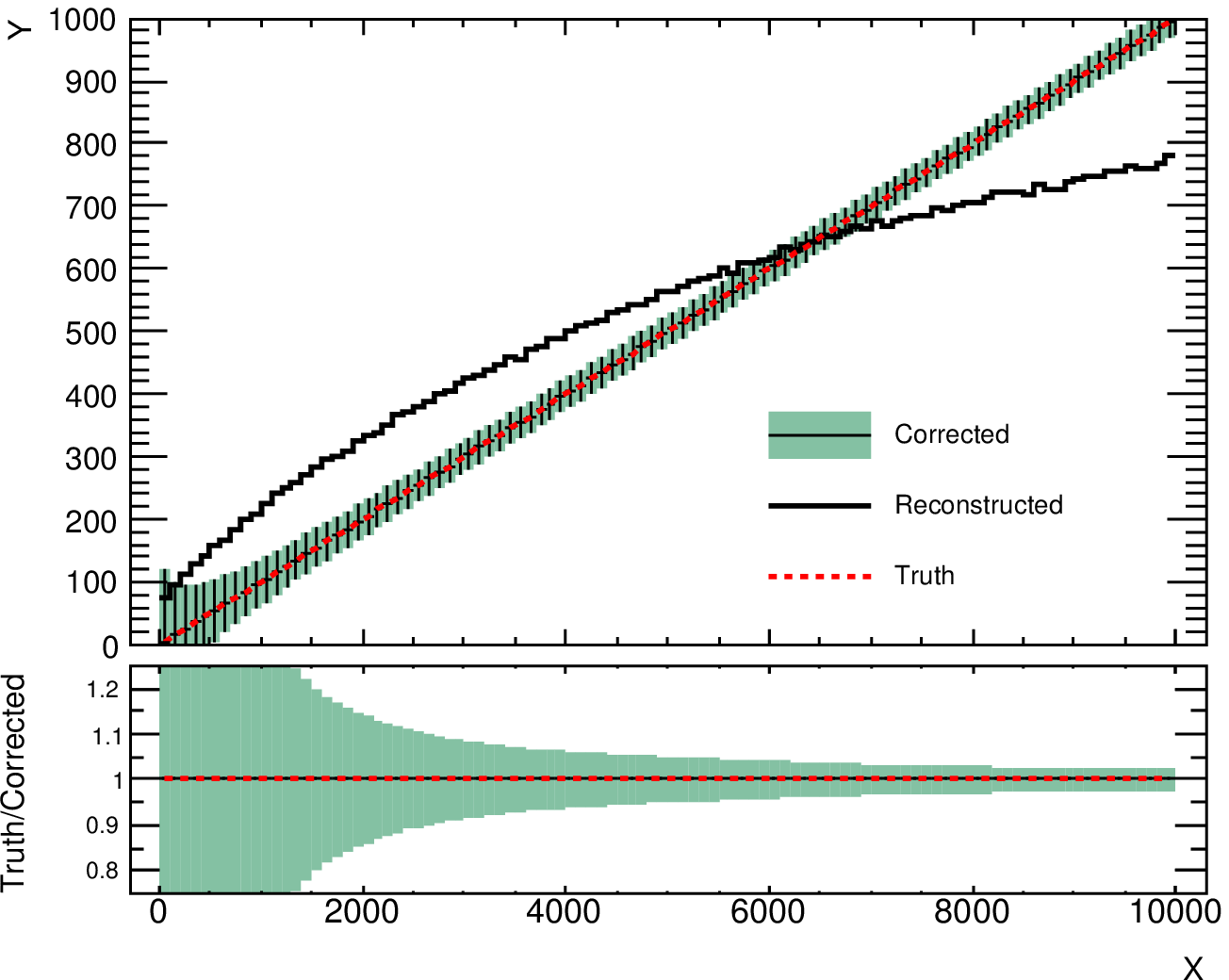}
        \caption[Two correlated variables are corrected for a Gaussian smearing, maintaining the correlation.]
        {Events have uniformly distributed values for the observables $X$ and $Y$, and $X=10Y$.
	A Gaussian smearing is applied, then Imagiro used to correct for this and recover the true variable distributions.
	In the upper plot, the points showing the corrected $X$ distribution fall exactly on the horizontal line of the true distribution.
	In the lower plot, the points showing the corrected $X$ vs $Y$ distribution fall exactly on the line $X=10Y$ of the true distribution.
	Both plots show perfect closure, despite each reconstructed distribution differing substantially from the truth.
	Corrected distributions show statistical errors as narrow bars, with a shaded background showing the statistical and systematic errors combined in quadrature.}
        \label{fig:LinearXvsY}
	\end{center}
\end{figure}

Two examples of results from unfolding with Imagiro are provided.
The first is for a single toy model, shown in Fig.~\ref{fig:LinearXvsY}.
Events are given uniform distributions in the arbitrary observables $X$ and $Y$, with $X$ ranging from $0$ to $10000$, and $Y$ from $0$ to $1000$.
The values of $X$ and $Y$ are correlated such that $X = 10Y$.
To create toy reconstructed distributions, a smearing is applied to each true value such that $X_\mathrm{reco} = X_\mathrm{true} ( 1 + q )$, and the same for $Y$, where $q$ is sampled each event from a Gaussian distribution of width $0.5$ and mean $0.0$.
Imagiro is then used to recover the true distributions of $X$ and $Y$.
The unfolding is perfect because the true distributions of $X$ and $Y$ are used as the prior distributions -- a closure test.
The correlation between $X$ and $Y$ is recovered using the method described in Section~\ref{subsec:UnfoldingCorrelated}.

\begin{figure}[tp]
	\begin{center}
	\includegraphics[width=0.7\textwidth]{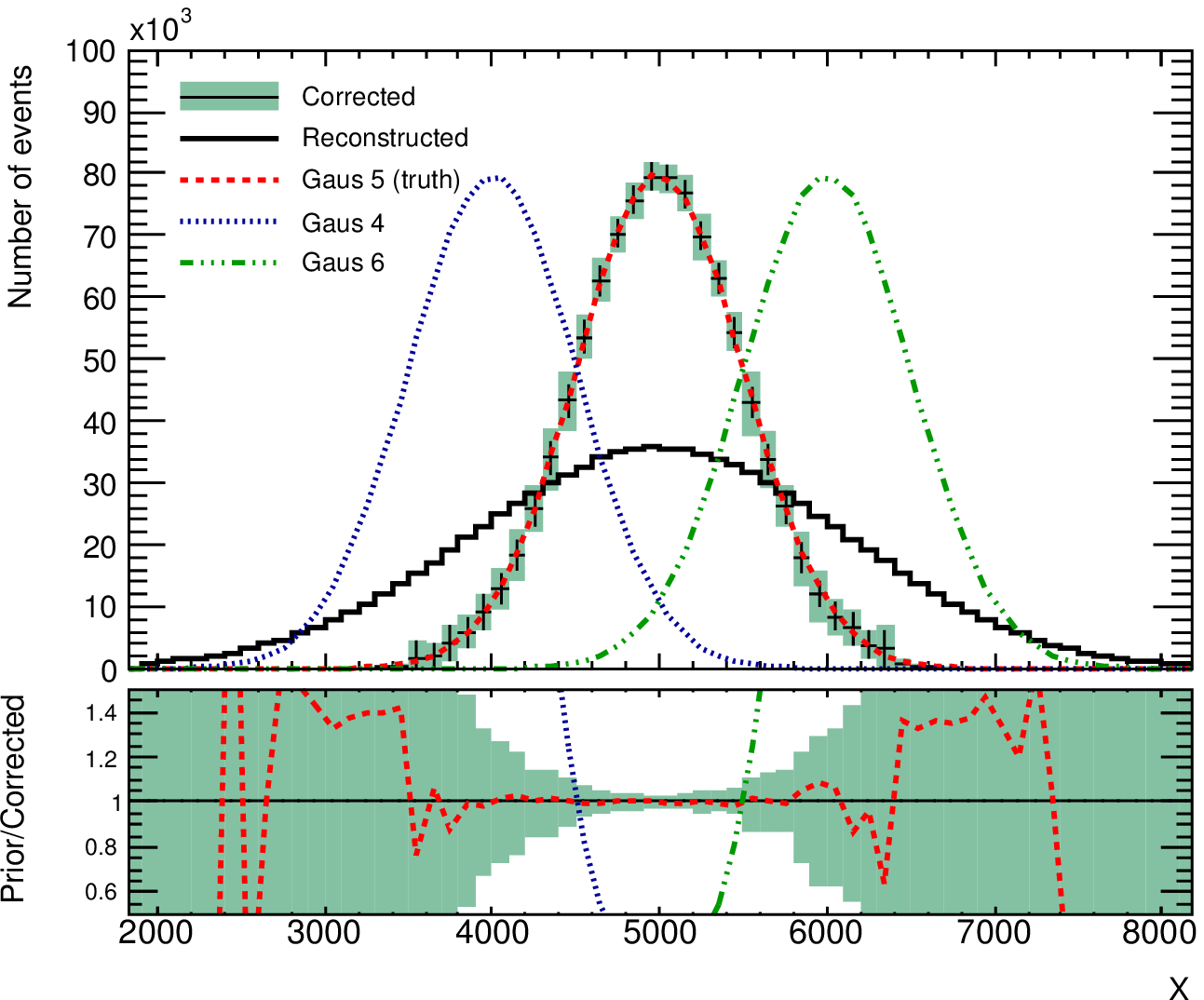}\\[1em]
	\includegraphics[width=0.7\textwidth]{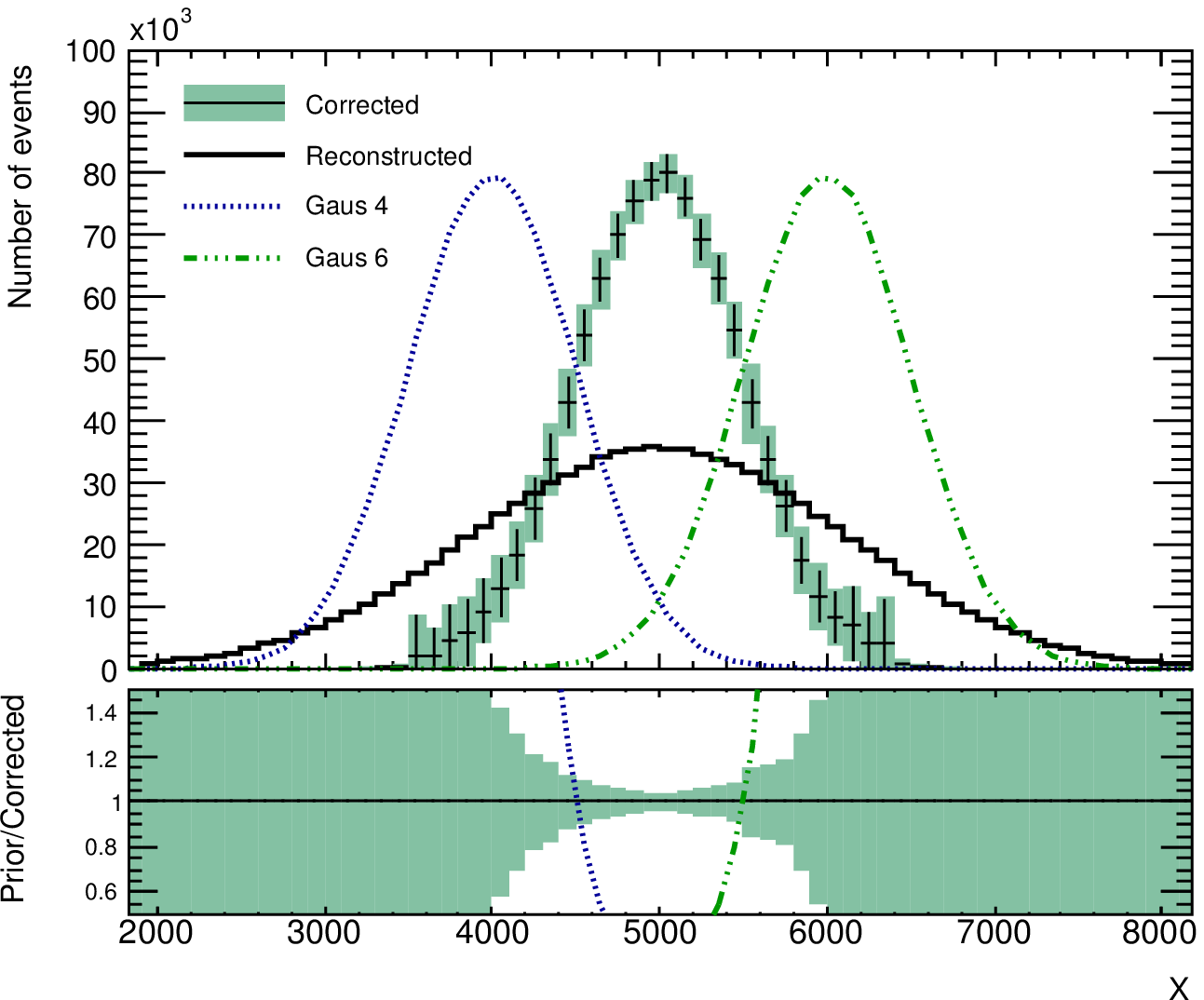}
	\caption[The distribution of a variable is corrected for a Gaussian smearing, using multiple different prior distributions]
	{In the upper plot, reconstructed events from Gaus~5 are unfolded using Gaus~4, Gaus~5 and Gaus~6 as priors.
	The corrected distribution closely matches Gaus~5.
	In the lower plot, reconstructed events from Gaus~5 are unfolded using only Gaus~4 and Gaus~6 as priors.
	Again, the corrected distribution closely matches Gaus~5, despite the fact that it was not used as a prior.
	Corrected distributions show statistical errors as narrow bars, with a shaded background showing the statistical and systematic errors combined in quadrature.}
	\label{fig:MultiplePriors}
	\end{center}
\end{figure}

In the second example, the observable $X$ has three different possible distributions.
Each is a Gaussian with width $500$, with a mean of either $4000$, $5000$, or $6000$.
These are referred to as `Gaus~4,' `Gaus~5,' and `Gaus~6' respectively.
To create toy reconstructed distributions a smearing is applied to each true value such that $X_\mathrm{reco} = X_\mathrm{true} + q$, where $q$ is sampled each event from a Gaussian distribution of width $1000$ and mean $0.0$.
The reconstructed events from Gaus~5 are then unfolded with Imagiro to recover their true distribution.
Each of the different possible distributions for $X$ is used as a prior for the unfolding, as shown in Fig.~\ref{fig:MultiplePriors}.
The original distribution for $X$ is recovered, with a systematic uncertainty given by the different results from unfolding with the different priors.
The unfolding is repeated using only Gaus~4 and Gaus~6 as priors (i.e.\ without using knowledge of the correct result) with almost identical results, although the statistical errors are larger because there are fewer events in the smearing matrix.

\section{User guide}

Imagiro is configured by editing two files: \InlineCode{main.cpp} and \InlineCode{Monte\-Carlo\-Information.cpp}.
Respectively, these define the observables to be unfolded and the MC samples available.
The software is then compiled and run without command line arguments or user interaction, producing a single output file in ROOT format.
A brief description of how to edit each of these files is given in this section.

Compiling Imagiro with the makefile provided requires a Linux or MacOS X operating system with ROOT installed.
The \InlineCode{ROOTSYS} and \InlineCode{LD\_LIBRARY\_PATH} environment variables must be correctly set.
Imagiro uses the \InlineCode{unordered\_map} class available in the C++11 standard, or earlier non-standard implementations such as in GCC 4.3.x.

\noindent
Imagiro is available from \relscale{0.9}{\url{http://cern.ch/imagiro}}.

\subsection{Defining Monte Carlo samples}

For each MC sample, Imagiro expects to load two instances of the ROOT class \InlineCode{TTree}, one for the truth events, and one for reconstructed.
These trees must contain branches called \InlineCode{Event\-Number} and \InlineCode{Event\-Weight}, where event number is a unique identifier to match corresponding truth and reconstructed events.
If an \InlineCode{Event\-Number} value from the truth tree cannot be found in the reconstructed tree that event is classed as missing, or fake if the reverse is true -- see Section~\ref{sec:FakeMiss}.
There should then be one branch for each observable to be unfolded, with corresponding branches named the same in each tree.
The trees can be in the same ROOT file, in separate files, or spread across multiple files.
In the simplest case of two trees in the same file \InlineCode{Monte\-Carlo\-Information.cpp} should contain an entry like this:
\begin{center}
\begin{tabular}{ p{0.9\textwidth} }
\hline
\vspace{-5px}
\begin{lstlisting}
combineFiles.push_back( false );
truthPaths.push_back( "path/to/file.root" );
recoPaths.push_back( "path/to/file.root" );
descriptions.push_back( "Description of sample" );
colours.push_back( kRed );
styles.push_back( 1 );
inputTypes.push_back( "InputNtuple" );
internalTruth.push_back( "truthTreeName" );
internalReco.push_back( "reconstructedTreeName" );
\end{lstlisting}
\vspace{-5px}\\
\hline
\end{tabular}
\end{center}
\InlineCode{Monte\-Carlo\-Information.cpp} populates each of these vectors, one entry per MC sample.
Their meanings are as follows:
\begin{itemize}
	\item \InlineCode{combine\-Files} contains \InlineCode{true} if each tree is spread across multiple files, or \InlineCode{false} otherwise.
	\item \InlineCode{truth\-Paths} contains the path to the ROOT file containing the MC truth tree.
	\item \InlineCode{reco\-Paths} contains the path to the ROOT file containing the reconstructed MC tree (in this case the same file).
	\item \InlineCode{descriptions} contains the text to describe this sample in plot legends.
	\item \InlineCode{colours} contains the ROOT colour to use for this sample in plots.
	\item \InlineCode{styles} contains the ROOT line style to use for this sample in plots.
	\item \InlineCode{input\-Types} contains the name of the class to use to load the \InlineCode{TTree}.
	\item \InlineCode{internal\-Truth} contains the path to, and name of the MC truth tree within the ROOT file.
	\item \InlineCode{internal\-Reco} contains the path to, and name of the reconstructed MC tree within the ROOT file.
\end{itemize}
If the MC truth and reconstructed trees are in separate files, change the \InlineCode{truth\-Paths} and \InlineCode{reco\-Paths} entries accordingly:
\begin{center}
\begin{tabular}{ p{0.9\textwidth} }
\hline
\vspace{-5px}
\begin{lstlisting}
...
truthPaths.push_back( "path/to/truthFile.root" );
recoPaths.push_back( "path/to/recoFile.root" );
...
\end{lstlisting}
\vspace{-5px}\\
\hline
\end{tabular}
\end{center}
When the trees are in separate files they are not required to have different names, so \InlineCode{internal\-Truth} and \InlineCode{internal\-Reco} can both contain the same value.

In cases where the truth and reconstructed trees are each spread across multiple files, three new vectors are used to describe them:
\begin{itemize}
	\item \InlineCode{extra\-Truth\-Paths} contains a vector of paths to additional root files containing MC truth trees.
	\item \InlineCode{extra\-Reco\-Paths} contains a vector of paths to additional root files containing reconstructed MC trees.
	\item \InlineCode{input\-Weights} contains a vector of weights for the events from each file.
\end{itemize}
Each of these is a vector of vectors so that many files can be combined at once.
For example:
\begin{center}
\begin{tabular}{ p{0.9\textwidth} }
\hline
\vspace{-5px}
\begin{lstlisting}
combineFiles.push_back( true );
truthPaths.push_back( "path/to/file.1.root" );
recoPaths.push_back( "path/to/file.1.root" );

vector< string > moreFiles;
moreFiles.push_back( "path/to/file.2.root" );
moreFiles.push_back( "path/to/file.3.root" );
extraTruthPaths.push_back( moreFiles );
extraRecoPaths.push_back( moreFiles );

vector< double > allFileWeights;
pythiaWeights.push_back( 1.0 );
pythiaWeights.push_back( 1.0 );
pythiaWeights.push_back( 1.0 );
inputWeights.push_back( pythiaWeights );
...
\end{lstlisting}
\vspace{-5px}\\
\hline
\end{tabular}
\end{center}
Note that it is assumed truth and reconstructed trees in the same file contain corresponding events.
If a truth event in \InlineCode{file.1.root} has the same \InlineCode{Event\-Number} as a reconstructed event in \InlineCode{file.2.root} they will not be matched, and will be treated as missing and fake respectively.
There must still be entries in \InlineCode{truth\-Paths} and \InlineCode{reco\-Paths}: it is the additional files that go into \InlineCode{extra\-Truth\-Paths} and \InlineCode{extra\-Reco\-Paths}.
The vector stored in \InlineCode{input\-Weights} must have one entry for every file to load, i.e.\ one more entry than in \InlineCode{extra\-Truth\-Paths} and \InlineCode{extra\-Reco\-Paths}.
This vector must be stored whenever trees are spread across multiple files, even if they should not have extra weighting applied -- use weights of \InlineCode{1.0} in that case.
All the other entries are as before, but \InlineCode{combine\-Files} should contain true in this case.
All tree parts must have the same format: branches cannot be added, removed or renamed between files.

The truth and reconstructed tree parts do not have to be in the same files.
The separate \InlineCode{extraTruthPaths} and \InlineCode{extraRecoPaths} vectors allow separate files to be used for each, as in this example:
\begin{center}
\begin{tabular}{ p{0.9\textwidth} }
\hline
\vspace{-5px}
\begin{lstlisting}
combineFiles.push_back( true );
truthPaths.push_back( "path/to/truthFile.1.root" );
recoPaths.push_back( "path/to/recoFile.1.root" );

vector< string > moreTruthFiles;
moreTruthFiles.push_back( "path/to/truthFile.2.root" );
moreTruthFiles.push_back( "path/to/truthFile.3.root" );
extraTruthPaths.push_back( moreTruthFiles );

vector< string > moreRecoFiles;
moreRecoFiles.push_back( "path/to/recoFile.2.root" );
moreRecoFiles.push_back( "path/to/recoFile.3.root" );
extraRecoPaths.push_back( moreRecoFiles );
...
\end{lstlisting}
\vspace{-5px}\\
\hline
\end{tabular}
\end{center}
The vectors must be ordered so that files containing corresponding events are at the same position in each vector.

The user can define multiple MC samples this way, with the condition that each one provides all of the observables to be unfolded, and that the branch for each observable is named the same in each sample.

\subsection{Requesting unfolded plots}

Each unfolded distribution of a variable is handled by an instance of the class \InlineCode{Monte\-Carlo\-Summary\-Plot\-Maker}.
These are all created in \InlineCode{main.cpp}, and should all be instantiated the same way:
\begin{center}
\begin{tabular}{ p{0.9\textwidth} }
\hline
\vspace{-5px}
\begin{lstlisting}
MonteCarloSummaryPlotMaker * myPlot = new MonteCarloSummaryPlotMaker(
                mcInfo, COMBINE_MC, globalRandom );
\end{lstlisting}
\vspace{-5px}\\
\hline
\end{tabular}
\end{center}
Here, \InlineCode{mcInfo} is a pointer to an instance of the \InlineCode{Monte\-Carlo\-Information} class that holds all the definitions of MC samples.
If systematic error studies are to be used then \InlineCode{global\-Random} should point to an instance of the ROOT random number generator \InlineCode{TRandom3}.
The boolean \InlineCode{COMBINE\_MC} is true if Imagiro should make a single smearing matrix from all MC samples, or false if there should be separate smearing matrices for each.

Now the plot should be defined.
For a single observable plot there are two ways of doing this, one for fixed histogram bin widths, and one for variable bins:
\begin{center}
\begin{tabular}{ p{0.9\textwidth} }
\hline
\vspace{-5px}
\begin{lstlisting}
myPlot->Make1DHistogram( "x", xBinNumber, xMin, xMax,
                PLOT_MODE, scaleFactor, normalise );
//or
myPlot->Make1DHistogram( "x", xBinLowEdges,
                PLOT_MODE, scaleFactor, normalise );
\end{lstlisting}
\vspace{-5px}\\
\hline
\end{tabular}
\end{center}
The first argument in either case is the name of the observable to unfold.
This should be identical to the name of the branch storing that observable in the input ROOT trees.
Then comes the definition of the histogram binning.
In the first case fixed width bins between \InlineCode{xMin} and \InlineCode{xMax} are created, dividing that range into \InlineCode{xBin\-Number} intervals.
In the second, the vector \InlineCode{xBin\-Low\-Edges} should contain one value indicating the lower edge of each bin to create, plus one value on the end to indicate the upper edge of the last bin.
The last three arguments have the same meaning in either case.
\InlineCode{PLOT\_MODE} affects the overall behaviour of Imagiro, and is discussed in Section~\ref{subsec:GlobalConstants}.
All bins in the plot will be multiplied by the value of \InlineCode{scale\-Factor}.
The histogram will be normalised if \InlineCode{normalise} is set to \InlineCode{true}.

For a profile showing the correlation between two unfolded observables there are again two possibilities:
\begin{center}
\begin{tabular}{ p{0.9\textwidth} }
\hline
\vspace{-5px}
\begin{lstlisting}
myPlot->MakeProfile( "x", "y", xBinNumber, xMin, xMax,
                yBinNumber, yMin, yMax,
                PLOT_MODE, scaleFactor );
//or
myPlot->MakeProfile( "x", "y", xBinLowEdges, yBinLowEdges,
                PLOT_MODE, scaleFactor );
\end{lstlisting}
\vspace{-5px}\\
\hline
\end{tabular}
\end{center}
Now there are two observable name arguments, and two sets of definitions for binning those observables, but in other respects everything is the same as for the single observable plot.
Note that \InlineCode{Make\-Profile} or \InlineCode{Make1D\-Histogram} can only be called once for each \InlineCode{Monte\-Carlo\-Summary\-Plot\-Maker}: each instance only handles a single plot.

There are additional configuration methods for the \InlineCode{Monte\-Carlo\-Summary\-Plot\-Maker} which change the appearance of the output plot:
\begin{center}
\begin{tabular}{ p{0.9\textwidth} }
\hline
\vspace{-5px}
\begin{lstlisting}
myPlot->SetYRange( plotMinimum, plotMaximum );
myPlot->SetAxisLabels( "Observable X", "Number of events" );
myPlot->UseLogScale();
\end{lstlisting}
\vspace{-5px}\\
\hline
\end{tabular}
\end{center}
\InlineCode{Set\-YRange} allows the user to define the visible region of the $y$-axis in the output plot.
\InlineCode{Set\-Axis\-Labels} gives two free text fields to label the $x$ and $y$ axes of the output plot.
If \InlineCode{Use\-Log\-Scale} is called then the plot will have a logarithmic $y$-axis.

Propagation of experimental systematic errors (as discussed in Section~\ref{subsec:SystematicPropagation}) is also set up through \InlineCode{Monte\-Carlo\-Summary\-Plot\-Maker} methods:
\begin{center}
\begin{tabular}{ p{0.9\textwidth} }
\hline
\vspace{-5px}
\begin{lstlisting}
myPlot->AddConstantSystematic( "x", systematicValue,
                numberOfPseudoExperiments, isAbsolute );

myPlot->AddPerEventSystematic( "x", "xSystematic",
                numberOfPseudoExperiments, isAbsolute );
\end{lstlisting}
\vspace{-5px}\\
\hline
\end{tabular}
\end{center}
When there is a single error value to be applied, use \InlineCode{Add\-Constant\-Systematic}.
The first argument is the observable to apply the error to, and the second is the size of that error.
To apply the error as a simple offset, use \InlineCode{number\-Of\-Pseudo\-Experiments = 1}.
For offsets at plus and minus the error value, use \InlineCode{number\-Of\-Pseudo\-Experiments = 2}.
Larger values for \InlineCode{number\-Of\-Pseudo\-Experiments} will instead cause the offset to be sampled each event from a Gaussian distribution with width given by the error value and mean \InlineCode{0.0}.
If the error should be treated as a fraction of the measured observable value, use \InlineCode{isAbsolute = false};

There is also the possibility to add an error value that changes each event, by loading it from the input data file.
In this case use \InlineCode{Add\-Per\-Event\-Systematic}, where the first argument is the name of the observable, and the second is the name of the branch containing the error value.
The other arguments behave in the same way as for the constant systematic.

Once the \InlineCode{Monte\-Carlo\-Summary\-PlotMaker} has been configured, it should be added to the vector of plots to process:
\begin{center}
\begin{tabular}{ p{0.9\textwidth} }
\hline
\vspace{-5px}
\begin{lstlisting}
allPlotMakers.push_back( myPlot );
\end{lstlisting}
\vspace{-5px}\\
\hline
\end{tabular}
\end{center} 

\subsection{Input data}

The source of data to unfold is also defined in \InlineCode{main.cpp}.
An instance of the abstract class \InlineCode{IFile\-Input} called \InlineCode{dataInput} must be created, as in the following example:
\begin{center}
\begin{tabular}{ p{0.9\textwidth} }
\hline
\vspace{-5px}
\begin{lstlisting}
IFileInput * dataInput = new InputNtuple( "path/to/data.root",
		"dataTreeName", "Description of data",
		mcInfo->NumberOfSources(), relevanceChecker );
\end{lstlisting}
\vspace{-5px}\\
\hline
\end{tabular}
\end{center}
The first three arguments are the path on disk to the ROOT file to load, the internal path and name of the \InlineCode{TTree} within the file, and the description of the data to use in plot legends, respectively.
Fourth must be an integer identifier for the data sample, which must be different to the (automatically defined) identifiers for the MC samples.
A useful value can be given by the instance of the \InlineCode{Monte\-Carlo\-Information} class, as shown.
Finally there is a pointer to an instance of the \InlineCode{Observable\-List} class, which ensures that only relevant information is loaded from the data file.

If the user wishes to try unfolding an already-defined MC sample for an explicit closure test, the information can be retrieved directly from the \InlineCode{Monte\-Carlo\-Information} class like this:
\begin{center}
\begin{tabular}{ p{0.9\textwidth} }
\hline
\vspace{-5px}
\begin{lstlisting}
IFileInput * dataInput = mcInfo->MakeTruthInput( 0, relevanceChecker );
//or
IFileInput * dataInput = mcInfo->MakeReconstructedInput(
		0, relevanceChecker );
\end{lstlisting}
\vspace{-5px}\\
\hline
\end{tabular}
\end{center}
The first argument is the index of the MC sample within \InlineCode{Monte\-Carlo\-Information}, and the second is the \InlineCode{Observable\-List} instance.

\subsection{Global constants}
\label{subsec:GlobalConstants}

Besides the definitions of plots to produce and data to unfold, \InlineCode{main.cpp} also contains global settings for Imagiro as constant values.
These are discussed below.

\noindent
\InlineCode{PLOT\_MODE} sets the correction to be applied to every plot (although it could be set differently for specific plots when their \InlineCode{Monte\-Carlo\-Summary\-Plot\-Maker} is configured).
It can take four possible integer values:
\begin{itemize}
	\item \InlineCode{2} indicates Bayesian iterative unfolding, as discussed in this paper.
	\item \InlineCode{1} indicates ``bin-by-bin'' unfolding: each bin in the measured data distribution is weighted by the ratio of the corresponding MC truth and reconstructed bins.
		This technique does not account for migrations of events between bins, and is model-dependent, so it is not recommended.
		It is provided for comparison.
	\item \InlineCode{0} will not perform any correction, inputs are plotted unchanged.
	\item \InlineCode{-1} indicates folding: the smearing matrix will be applied to the input distribution as in Eqn.~\ref{eqn:ApplySmearing}.
		This may be useful to quickly produce detector-level distributions from MC truth, without requiring the detector simulation to be performed.
\end{itemize}

\noindent
\InlineCode{COMBINE\_MC} sets whether to use a single smearing matrix from all MC samples, or to use a separate matrix for each one, as discussed in Section~\ref{subsec:UnfoldingPriors}.
This could be set separately for each plot when their \InlineCode{Monte\-Carlo\-Summary\-Plot\-Maker} is instantiated.

\noindent
\InlineCode{ERROR\_MODE} sets the method to use in calculating the statistical errors.
It can take three possible integer values:
\begin{itemize}
	\item \InlineCode{2} indicates the full statistical error calculation, as described in Section~\ref{subsec:StatisticalErrors}.
		This is only valid when using \InlineCode{PLOT\_MODE = 2}.
	\item \InlineCode{1} is like \InlineCode{2}, but only the diagonal elements of the covariance matrix are calculated, in order to save time.
		This is only valid when using \InlineCode{PLOT\_MODE = 2}.
	\item \InlineCode{0} indicates fast statistical error calculation, as described in Section~\ref{subsec:FastErrors}.
\end{itemize}

\noindent
\InlineCode{ITERATION\_NUMBER} is an unsigned integer setting the number of iterations to use in Bayesian unfolding.
To use automatic determination of the optimum iteration number (as described in Section~\ref{subsec:UnfoldingConvergence}) set it to \InlineCode{0}.
If setting the number of iterations manually, between \InlineCode{3} and \InlineCode{5} is recommended.

\noindent
\InlineCode{WITH\_SMOOTHING} sets whether to smooth the prior distribution each iteration before unfolding with it.
Currently this gives poor results -- possibly due to the unsophisticated smoothing algorithm -- and is recommended to be set to false.

\noindent
\InlineCode{OUTPUT\_FILE\_NAME} is the path and file name for the output ROOT file.

\section{Conclusion}

Imagiro is new software to correct data distributions for detector effects, using Bayesian iterative unfolding.
The algorithm is implemented with a full statistical error calculation.
Experimental systematic errors are propagated using pseudo-experiments, and the systematic error intrinsic to the correction is estimated using multiple prior distributions.
Cross-checking between different prior distributions allows the ideal number of unfolding iterations to be calculated.
Closure tests are performed automatically to test the validity of the correction.
Correlation between observables can be preserved through unfolding with the construction of a combined observable.
Imagiro is currently being used for several analyses in the ATLAS soft-QCD group.

\acknowledgments

I would like to thank Andy Buckley for advice when writing Imagiro, Tim Adye for discussion of the statistical error treatment, and also Andy Buckley (again), Phil Clark and Andreas Sch\"{a}like for their help editing this document.
I am principally supported by a PhD studentship from the UK Engineering and Physical Sciences Research Council.

\bibliographystyle{JHEP}
\bibliography{imagiro.bib}

\end{document}